\documentclass[sigconf, screen]{acmart}

\AtBeginDocument{%
  }

\copyrightyear{2026}
\acmYear{2026}
\setcopyright{cc}
\setcctype{by}
\acmConference[DAC '26]{63rd ACM/IEEE Design Automation Conference}{July 26--29, 2026}{Long Beach, CA, USA}
\acmBooktitle{63rd ACM/IEEE Design Automation Conference (DAC '26), July 26--29, 2026, Long Beach, CA, USA}
\acmDOI{10.1145/3770743.3803899}
\acmISBN{979-8-4007-2254-7/2026/07}

\usepackage{makecell}
\usepackage{bm}
\usepackage{subfig}
\usepackage{multirow}
\usepackage{multicol}
\usepackage{booktabs}
\usepackage[ruled,linesnumbered]{algorithm2e}
\usepackage{makecell}
\usepackage{soul}
\usepackage[most]{tcolorbox}
\usepackage{pifont}
\usepackage{multirow}

\usepackage{diagbox}

\usepackage[most]{tcolorbox} 

\usepackage{url}

\definecolor{MyPurple}{HTML}{58429B}
\definecolor{MyBrown}{HTML}{994c00}

\definecolor{MyDeepRed}{HTML}{990000} % dark
\definecolor{MyDeepBlue}{HTML}{0066cc} % light
\definecolor{MyColorGreen}{HTML}{a7c682}
\definecolor{MyColorBlue}{HTML}{49519a}

\begin{document}

\title{GEMM-GS: Accelerating 3D Gaussian Splatting on Tensor Cores with GEMM-Compatible Blending}

\author{
Haomin Li$^{1,\dag}$, Bowen Zhu$^{1,\dag}$, Fangxin Liu$^{1,2,\dag,\ddag}$, Zongwu Wang$^{1,2}$, Xinran Liang$^{3}$, Li Jiang$^{1,2,\ddag}$, and Haibing Guan$^{1}$}
\thanks{$\dag$ These authors contributed equally to the paper. $\ddag$ Corresponding authors.}
% \thanks{}
\thanks{This work was partially supported by the National Key Research and Development Program of China (2024YFE0204300), National Natural Science Foundation of China (Grant No.62402311), Natural Science Foundation of Shanghai (Grant No.24ZR1433700), and Key Research and Development Program of Shanghai (25LN3201200).}

\affiliation{%
  \institution{1.Shanghai Jiao Tong University, \,\, 2. Shanghai Qi Zhi Institute, \,\, 3. United Imaging Intelligence Co., Ltd.}
  \{haominli,\, liufangxin, \, ljiang\_cs\}@sjtu.edu.cn
  \country{}}

\renewcommand{\shortauthors}{Haomin Li, Bowen Zhu, Fangxin Liu, et al.}

\begin{abstract}
Neural Radiance Fields (NeRF) enables 3D scene reconstruction from several 2D images but incurs high rendering latency via its point-sampling design. 
3D Gaussian Splatting (3DGS) improves on NeRF with explicit scene representation and an optimized pipeline yet still fails to meet practical real-time demands. 
Existing acceleration works overlook the evolving Tensor Cores of modern GPUs because 3DGS pipeline lacks General Matrix Multiplication (GEMM) operations.
This paper proposes GEMM-GS, an acceleration approach utilizing tensor cores on GPUs via GEMM-friendly blending transformation.
It equivalently reformulates the 3DGS blending process into a GEMM-compatible form to utilize Tensor Cores. 
A high-performance CUDA kernel is designed, integrating a three-stage double-buffered pipeline that overlaps computation and memory access. 
Extensive experiments show that GEMM-GS achieves $1.42\times$ speedup over vanilla 3DGS and provides an additional $1.47\times$ speedup on average when combining with existing acceleration approaches.
Code is released at \url{https://github.com/shieldforever/GEMM-GS}.
\end{abstract}

% \keywords{Neural Rendering, 3D Gaussian Splatting (3DGS), GPU Acceleration, Tensor Core, General Matrix Multiplication (GEMM)}

\maketitle

\section{Introduction}

Neural rendering~\cite{nerf} has emerged as a transformative approach in bridging computer graphics and vision, leveraging differentiable rendering to reconstruct photo-realistic images and videos with exceptional fidelity. Early neural rendering techniques, such as Neural Radiance Fields (NeRF), marked a significant departure from traditional methods reliant on 3D grids and voxels. By harnessing the generalization capabilities of neural networks, NeRF implicitly reconstructs 3D scenes from a sparse set of 2D images, enabling high-quality rendering from novel viewpoints. This makes NeRF particularly valuable for applications like augmented reality (AR), virtual reality (VR), and 3D reconstruction~\cite{Zhou_2023_CVPR,wang2022nerf,10.1145/3528233.3530718,tancik2022block}. However, NeRF's sampling-point-based design requires over a million model inferences per image, resulting in significant rendering latency that hinders real-time performance.

Recently, 3D Gaussian Splatting (3DGS)~\cite{kerbl20233d} has gained prominence as a highly efficient neural rendering technique.
3DGS models the scenes with parametric point clouds of differentiable 3D Gaussian primitives.
By employing explicit scene representations and a well-designed rendering pipeline, 3DGS delivers better rendering quality and lower latency compared to NeRF-based methods.
Therefore, 3DGS has emerged as a leading method for 3D reconstruction and differentiable rendering, widely adopted across the field.

Despite its advancements, 3DGS falls short of real-time performance required for practical applications. 
To be specific, as the scale and resolution of images increase, both the number of Gaussians and the size of each Gaussian grow. 
At the hardware level, some domain specific architectures (DSAs) have been proposed to accelerate the rendering process, but these involve significant design and tape-out costs~\cite{lee2024gscore,li2026orange,liu2025asdr,lin2025metasapiens,feng2025lumina,wu2024gauspu}. 
In addition, the DSAs are hard to adapt to various optimization approaches.
At the software level, some works~\cite{fan2024lightgaussian,girish2024eagles,lee2024compact,papantonakis2024reducing} adopt mature quantization or pruning methods to reduce storage cost.
For example, LightGaussian~\cite{fan2024lightgaussian} adopts Gaussian pruning to reduce the huge number of Gaussians in the models and vector quantization to compress the attributes of Gaussians.
However, these methods mainly focus on storage reduction, but have not yielded significant performance improvements. 
Some other software efforts~\cite{feng2025flashgs,hanson2025speedy,radl2024stopthepop} focus on optimizing GPU implementation, achieving performance gains by improving the precision of intersection detection during the preprocessing stage to reduce redundant duplication of Gaussians.

\begin{figure}[tp] 
    \setlength{\abovecaptionskip}{1pt}  
    \setlength{\belowcaptionskip}{1pt} 
    \centering
    \includegraphics[width=1\linewidth]{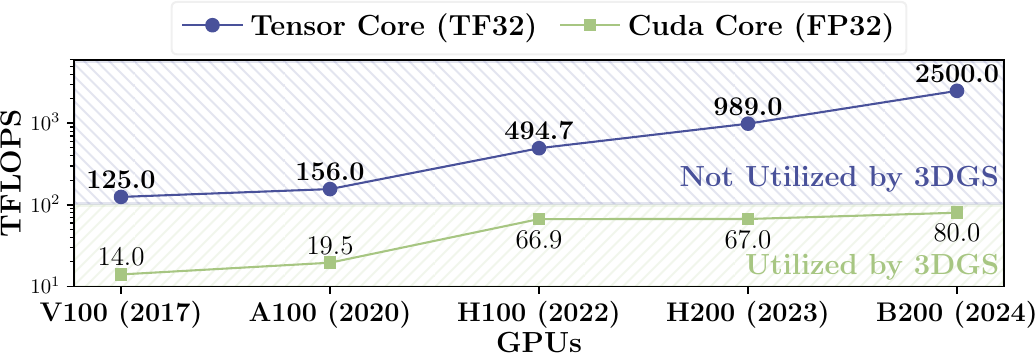}
    \caption{ 
    Computing Power Breakdown of modern GPUs utilized by 3D Gaussian Splatting. Data is collected from the data sheets of the GPU products~\cite{V100_DataSheet,A100_DataSheet,H100_DataSheet,H200_DataSheet,B200_DataSheet}.
    }
    \label{exp-fig:intro-gpu}
    \vspace{-0.6cm}
\end{figure}

\begin{figure*}[tp] 
    \setlength{\abovecaptionskip}{1pt}  
    \setlength{\belowcaptionskip}{1pt} 
    \centering
    \includegraphics[width=0.95\linewidth]{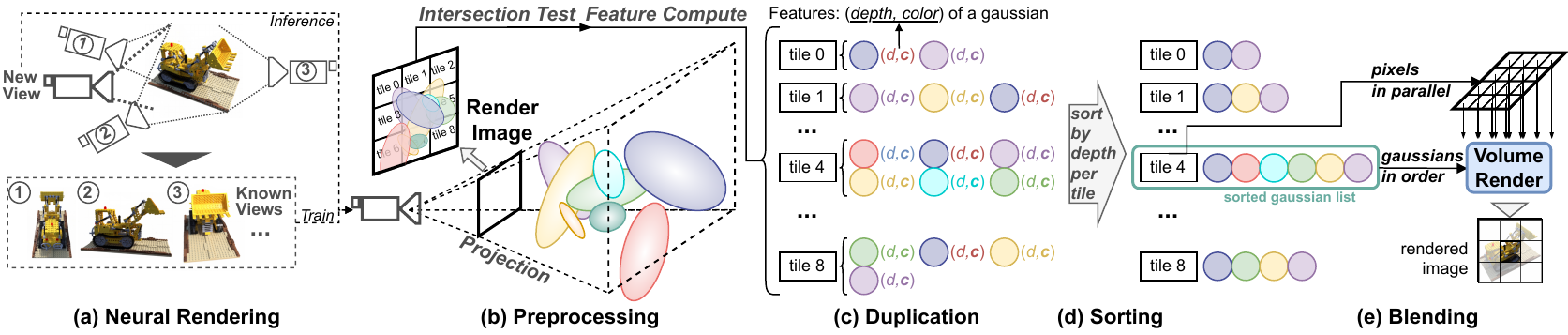}
    \caption{ Neural Rendering and Process of 3DGS~\cite{kerbl20233d}. 
        (a) Neural rendering (process of novel view synthesis). 3DGS consists of three stages. 
        (b) Stage 1: Preprocessing. Gaussians are projected onto the render image and intersection test is performed to relating projected Gaussians and tiles. Gaussians' features are also computed, including depth $d$ and color $\mathbf{c}$. 
        (c) Stage 2: Duplication. Each Gaussian is duplicated according to the number of tiles it intersects with.
        (d) Stage 3: Sorting. Gaussians in each tile are sorted by depth $d$.
        (e) Stage 4: Blending. The pixels in one tile are rendered by volume rendering in parallel, with the same sorted Gaussian list.
    }
    \label{fig:background-3dgs}
    \vspace{-0.4cm}
\end{figure*}

However, existing software optimization methods overlook the evolutionary characteristics of modern GPUs, namely the increasingly powerful TensorCores capable of efficiently performing general matrix multiplication (GEMM). 
That's because the rendering pipeline of 3DGS contains few GEMM-style operations, resulting in idle Tensor Cores on GPUs.
As shown in Figure~\ref{exp-fig: intro gpu}, starting from the Volta architecture, the computing power of GPU Tensor Cores has become increasingly stronger. Therefore, how to design a GEMM-friendly 3DGS rendering pipeline to fully utilize the powerful computing capabilities of Tensor Cores is the key to improving 3DGS performance.

To this end, this paper proposes GEMM-GS, an acceleration approach utilizing tensor cores on GPUs via GEMM-friendly blending transformation.
We transform the conventional 3DGS blending process into a GEMM-style formulation equivalently, leveraging the powerful tensor cores in modern GPUs for acceleration.
To fully utilize the computing resources, we develop a well-designed blending kernel in CUDA.
Based on the asynchronous memory copy technique, the kernel is designed as a three-stage pipeline with double buffer, well overlapping the computation and memory access.
The results demonstrate that GEMM-GS delivers $1.42\times$ speedup over the vanilla 3DGS implementation on average.
It is worth noting that our approach is orthogonal to existing acceleration methods, and can achieve an additional $1.47\times$ speedup over them.
Our contributions are summarized as follows:
\begin{itemize}
    \item We propose an equivalent transformation to reformulate the 3DGS blending process into a GEMM-compatible form, enabling the effective utilization of idle Tensor Cores on GPUs.
    \item We develop a CUDA kernel to implement the transformed blending process. The kernel adopts a three-stage double-buffered pipeline with asynchronous memory copy, fully overlapping Tensor Core computation and memory access.
    \item Evaluations demonstrate that GEMM-GS achieves an average $1.42\times$ speedup over vanilla 3DGS and an additional $1.47\times$ speedup when combined with existing methods. Moreover, GEMM-GS is directly deployable on commodity GPUs without hardware modification, ensuring low adoption cost and real-world applicability.
\end{itemize}

\section{Background}

\subsection{3D Gaussian Splatting}

3D Gaussian Splatting (3DGS) represents a scene using Gaussian ellipsoids and renders it in a tile-by-tile fashion. As illustrated in Figure~\ref{fig:background-3dgs}, the rendering pipeline consists of four stages: preprocessing, duplication, sorting, and blending.
\textbf{(a) Preprocessing:}  
Given a camera pose, 3D Gaussians within the view frustum are projected onto the 2D image plane (from ellipsoids to ellipses). The 2D screen is divided into tiles (e.g., $16 \times 16$ pixels) based on the target resolution and tile size. Intersection detection identifies which Gaussians contribute to each tile. For each projected Gaussian, depth $d$ and RGB color $\mathbf{c}$ are computed for use in subsequent stages.
\textbf{(b) Duplication:}  
To facilitate parallel blending, each Gaussian is duplicated according to the number of tiles it overlaps. Specifically, each copy is assigned a tile index, which is concatenated with the depth value to ensure that Gaussians corresponding to the same tile are gathered after sorting.
\textbf{(c) Sorting:}  
Each tile may intersect multiple Gaussians, with closer Gaussians occluding those farther away. To ensure correct rendering, Gaussians are sorted by depth $d$. GPU implementations typically employ radix sort~\cite{kerbl20233d} to exploit CUDA core parallelism, producing a sorted list of Gaussians for each tile.
\textbf{(d) Blending:}  
Blending iterates through the sorted Gaussians for each tile, processing all pixels in parallel. For each Gaussian $\mathcal{G}_i$ and pixel $p_j$, the opacity $\alpha_{ij}$ and the corresponding color contribution to $p_j$ are computed according to standard volume rendering operations:
\begin{equation}
\label{eq:volume-render}
\mathbf{C}_j = \sum_{i=1}^{N} T_i \alpha_{ij} \mathbf{c}_i, \quad
T_i = \prod_{k=1}^{i-1} (1 - \alpha_{kj}),
\end{equation}
where $\mathbf{C}_j$ is the rendered color of pixel $p_j$, and $T_i$ represents the accumulated transmittance up to Gaussian $i$.

As shown in Algorithm~\ref{alg:vanilla-blend}, rendering a $16 \times 16$ tile is assigned to a thread block of 256 threads, with each thread handling one pixel. For Gaussians stored in global memory, the thread block processes them in batches, loading one batch into shared memory at a time. Each thread iterates over the Gaussian attributes to compute the pixel color according to Eq.(\ref{eq:volume-render}). This tile-based organization enables coalesced memory accesses and efficient use of shared memory, reducing memory bandwidth pressure and improving throughput.

% 原始blending伪代码
\begin{algorithm}[tp]
    \caption{Blending Process in Vanilla 3DGS~\cite{kerbl20233d}.}
    \label{alg:vanilla-blend}
    
    \let\oldnl\nl% Store \nl in \oldnl
    \newcommand{\nonl}{\renewcommand{\nl}{\let\nl\oldnl}}% Remove line number for one line
    
    \KwData{
    Index list of sorted projected Gaussians $\mathcal{G}$ overlapping with current tile, batch size $b$
    }
    \KwResult{Rendered Color $\textbf{C}_j$s of pixels in current tile}

    \For{\textnormal{each batch ($b$ Gaussians)} $\subset \mathcal{G}$}
    {   

        \textcolor{MyColorBlue}{/************ \texttt{\textbf{Prepare Data}} *************/}

        \For{\textnormal{each} Gaussian $\mathcal{G}_i$ in current batch $\textnormal{\textbf{\underline{parallel}}}$ }
        {
            \textbf{Fetch} data of Gaussian $\mathcal{G}_i$: 

            \ \ \ \ projected coordinates: ($x_{g_i}$, $y_{g_i}$)
            
            \ \ \ \ covariance matrix: $\Sigma_i^{-1}=\begin{bmatrix}
                A_i & B_i\\
                B_i & C_i
                \end{bmatrix}$

            \ \ \ \ opacity weight and RGB color: $o_i$ and $\textbf{c}_i$
        }
        
        \textcolor{MyColorBlue}{/************ \texttt{\textbf{Volume Render}} ************/}

        \For{\textnormal{each} Gaussian $\mathcal{G}_i$ in current batch}
        {

            \For{\textnormal{each pixel} $p_j$ \textnormal{in current tile} \textnormal{\textbf{\underline{parallel}}}}
            {
                $\textbf{x}_{g_i} \leftarrow [x_{g_i}-x_{p_j}\ \ y_{g_i}-y_{p_j}]^T$

                $\alpha_{ij} \leftarrow o_i \times exp( {-\frac{1}{2}\textbf{x}_{g_i}^T\Sigma_i^{-1}\textbf{x}_{g_i}} ) $

                \If{$\alpha_{ij} \leq \frac{1}{255}$}
                {
                    \textnormal{\textbf{skip} to next Gaussian}
                    \hfill
                    \textcolor[RGB]{88, 100, 113}{// $\alpha$-skipping \cite{kerbl20233d}}
                }

                $T_{j} \leftarrow T_{j} \times (1-\alpha_{ij})$

                \If{$T_{j} \leq 0$}
                {
                    \textnormal{\textbf{stop} rendering $p_j$}
                    \hfill
                    \textcolor[RGB]{88, 100, 113}{// early terminate\cite{nerf}}
                }

                $\textbf{C}_{j} \leftarrow \textbf{C}_{j} + \textbf{c}_{i}\alpha_{ij}T_{j}$
            }
        }
    }
\end{algorithm}

\subsection{3DGS GPU Acceleration}

To improve the rendering speed of 3DGS models, existing acceleration approaches can be broadly categorized into compression-based and preprocessing-based methods.

\textbf{Compression-based approaches} aim to reduce the memory footprint of Gaussian representations. Two common techniques are pruning, which reduces the number of Gaussians based on importance metrics~\cite{Hanson_2025_CVPR,mallick2024taming}, and vector quantization (VQ), which represents the attribute vectors of a large number of Gaussians using a small set of cluster centroids~\cite{navaneet2023compact3d,niedermayr2024compressed,Dai_2025_CVPR}. Some works, such as LightGaussian~\cite{fan2024lightgaussian}, combine pruning and VQ to achieve further storage reduction~\cite{lee2024compact,papantonakis2024reducing,girish2024eagles}. However, these approaches typically require retraining the model to restore rendering quality.

\textbf{Preprocessing-based approaches} focus on optimizing Gaussian intersection tests to avoid redundant processing, without modifying the underlying 3DGS models. FlashGS~\cite{feng2025flashgs} employs a precise redundancy elimination algorithm with opacity skipping. StopThePop~\cite{radl2024stopthepop} introduces a tile-based culling strategy similar to FlashGS. Speedy-Splat~\cite{hanson2025speedy} proposes the SnugBox algorithm, which constructs tight bounding boxes around each Gaussian and identifies the exact set of intersected tiles. These methods are lossless and do not require retraining or model modification.

\subsection{Analysis and Motivation}
\begin{figure}[tp] 
    \setlength{\abovecaptionskip}{3pt}  
    \setlength{\belowcaptionskip}{3pt} 
    \centering
    \includegraphics[width=1\linewidth]{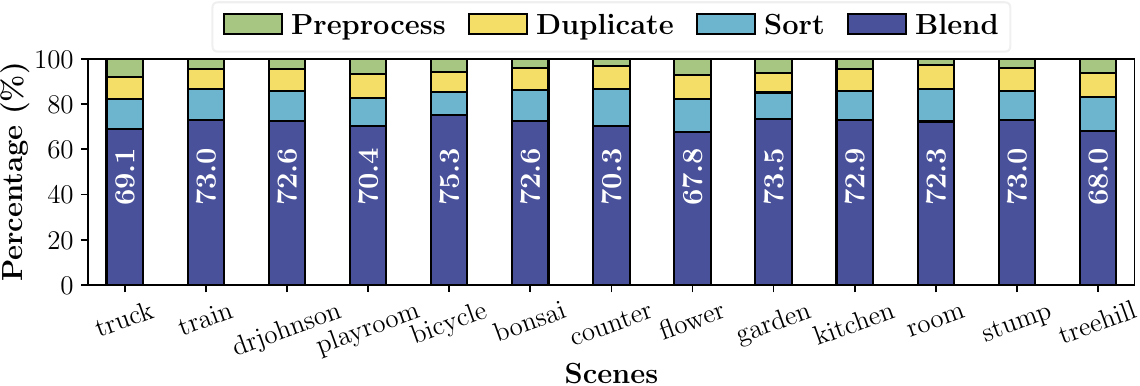}
    \caption{ Rendering Latency Breakdown of 3DGS. The scenes for rendering are from three datasets: Tank\&Temples~\cite{knapitsch2017tanks}, Deep Blending~\cite{hedman2018deep}, and Mip-NeRF 360~\cite{barron2022mip}.}
    \label{exp-fig: motivation breakdown profiling}
    \vspace{-0.3cm}
\end{figure}

We first profile 3DGS rendering latency across scenes from three datasets~\cite{knapitsch2017tanks,hedman2018deep,barron2022mip} on an NVIDIA A100 GPU (Figure~\ref{exp-fig: motivation breakdown profiling}). 
The results show that the blending stage contributes nearly $70\%$ of the total rendering time, making it the primary performance bottleneck. 
At the same time, modern GPUs feature increasingly powerful Tensor Cores optimized for GEMM-based computations. As illustrated in Figure~\ref{exp-fig:intro-gpu}, the theoretical FLOPS of Tensor Cores on the latest GPUs can exceed 30 times that of CUDA cores. However, the conventional 3DGS pipeline contains very few GEMM-style operations, leaving these resources largely idle and representing a substantial opportunity for acceleration.

These observations motivate our primary optimization target: transforming the blending stage into a GEMM-compatible pipeline to leverage the idle Tensor Cores. To this end, we propose \textbf{GEMM-GS}, which constructs a GEMM-compatible computation pipeline by applying an equivalent transformation to the blending stage and implements a high-performance CUDA kernel for accelerated Gaussian blending. Our approach is orthogonal to existing GPU acceleration methods and provides a plug-and-play tensor-based computation framework, enabling further acceleration of various state-of-the-art 3DGS implementations.

\section{GEMM-GS Design and Implementation}
The blending stage dominates the computational cost of 3DGS. To exploit Tensor Cores, GEMM-GS reformulates the opacity computation ($\alpha$) into a GEMM-compatible form and designs a high-performance CUDA kernel that fully leverages the GPU memory hierarchy and parallel execution resources.
\subsection{Calculation Form of $\alpha$'s Exponential}

In 3DGS, the opacity $\alpha_{ij}$ of a Gaussian $\mathcal{G}i$ at pixel $p_j$ is computed as:
\begin{equation}
\label{eq: alpha}
\begin{aligned}
    \alpha_{ij}= o_i \times e^{-\frac{1}{2}\textbf{x}_{g_i}^T\Sigma_i^{-1}\textbf{x}_{g_i}}
\end{aligned}
\end{equation}
where $\mathbf{x}_{g_i}=[\Delta x_{ij}\ \ \Delta y_{ij}]^T=[x_{g_i}-x_{p_j}\ \ y_{g_i}-y_{p_j}]^T$, i.e., the vector from the projected coordinate of Gaussian $g_i$ $(x_{g_i}, y_{g_i})$ to the coordinate of the current pixel $(x_{p_j}, y_{p_j})$. $\Sigma_i^{-1}=\begin{bmatrix}
A_i & B_i\\
B_i & C_i
\end{bmatrix}$ is the covariance matrix of $g_i$.

Thus, the exponential term in Eq.(\ref{eq: alpha}) is expanded as
\begin{equation}
\label{eq: power}
\begin{aligned}
    power_{ij}
    &=-\frac{1}{2}A_i\Delta x_{ij}^2 - B_i\Delta x_{ij}\Delta y_{ij} - \frac{1}{2}C_i\Delta y_{ij}^2
\end{aligned}
\end{equation}
This formulation involves element-wise quadratic terms, which cannot be directly expressed as a matrix multiplication. As a result, the blending stage cannot natively benefit from Tensor Cores, leading to low compute efficiency on modern GPUs. GEMM-GS addresses this by transforming the computation into a GEMM-compatible representation, enabling effective Tensor Core acceleration.

\subsection{Transformation based on Intra-Tile Coordinates}
\label{sec:transf}
To enable Tensor Core utilization, we reformulate the opacity computation in the blending stage into a dot-product form. The key idea is to exploit intra-tile relative coordinates so that the quadratic power term can be expressed as a linear dot product, which is then amenable to batched matrix multiplications. For each tile, we select a reference pixel $p_c$ (e.g., the center pixel) with coordinates $(x_{c}, y_{c})$. The coordinates of any pixel $p_j$ in the tile can be expressed as:
\begin{equation}
(x_{p_j}, y_{p_j}) = (x_{c} - \bar{x}_{p_j}, \ y{c} - \bar{y}_{p_j}),
\end{equation}
where $(\bar{x}{p_j}, \bar{y}{p_j})$ are the intra-tile relative coordinates of $p_j$ with respect to $p_c$.
Accordingly, the coordinate differences between a Gaussian $\mathcal{G}i$ and pixel $p_j$ become:
\begin{equation}
    \label{eq: delta rewritten}
    \begin{aligned}
        \Delta x_{ij}
        =\textcolor{MyColorBlue}{(x_{g_i}-x_{c})}+\bar{x}_{p_j}
        ,\ 
        \Delta y_{ij}
        =\textcolor{MyColorBlue}{(y_{g_i}-y_{c})}+\bar{y}_{p_j}
    \end{aligned}
\end{equation}
We let $\hat{x}_{g_i}$ and $\hat{y}_{g_i}$ to denote the \textcolor{MyColorBlue}{blue parts} $x_{g_i} - x_{c}$ and $ = y_{g_i} - y_{c}$, which are the coordinate differences between Gaussian $\mathcal{G}_i$ and the reference pixel $p_c$. 
These pairs of values ($\hat{x}_{g_i}$, $\hat{y}_{g_i}$)s are constant for all pixels in a tile for a given Gaussian, requiring computation only once per tile.

Substituting these terms into Eq.(\ref{eq: power}), the quadratic form can be reorganized as:
\begin{equation}
\begin{aligned}
    \label{eq: power rewritten}
    power_{ij}=&-\frac{1}{2}A_i(\hat{x}_{g_i}+\bar{x}_{p_j})^2 - B_i(\hat{x}_{g_i}+\bar{x}_{p_j})(\hat{y}_{g_i}+\bar{y}_{p_j}) 
    \\
    &- \frac{1}{2}C_i(\hat{y}_{g_i}+\bar{y}_{p_j})^2\\
    =&
    \textcolor{MyDeepRed}{
    \begin{bmatrix}
        -\frac{1}{2}A_i \\ 
        -\frac{1}{2}C_i \\ 
        -B_i \\ 
        -A_i \hat{x}_{g_i}-B_i \hat{y}_{g_i} \\ 
        -C_i \hat{y}_{g_i}-B_i \hat{x}_{g_i} \\ 
        -\frac{1}{2}A_i \hat{x}_{g_i}^2-\frac{1}{2}C_i \hat{y}_{g_i}^2-B_i \hat{x}_{g_i}\hat{y}_{g_i} \\
    \end{bmatrix}^T
    }
    \cdot
    \textcolor{MyColorBlue}{
        \begin{bmatrix}
            {\bar{x}_{p_j}}^2 \\
            {\bar{y}_{p_j}}^2 \\
            \bar{x}_{p_j}\bar{y}_{p_j} \\
            \bar{x}_{p_j} \\
            \bar{y}_{p_j} \\
            1 \\
        \end{bmatrix}
    }\\
    =\ \textcolor{MyDeepRed}{\Vec{\bf{v_{g_i}}}} \cdot \textcolor{MyColorBlue}{\Vec{\bf{v_{p_j}}}} 
\end{aligned}
\end{equation}
where $\textcolor{MyDeepRed}{\Vec{\bf{v_{g_i}}}}\in \mathbb{R}^6$ is a vector determined by Gaussian parameters ($A_i, B_i, C_i$) and offsets $(\hat{x}_{g_i}, \hat{y}_{g_i})$, and $\textcolor{MyColorBlue}{\Vec{\bf{v_{p_j}}}}\in \mathbb{R}^6$ encodes intra-tile relative coordinate terms for pixel $p_j$.
Notably, the \textcolor{MyColorBlue}{$\Vec{\bf{v_{p_j}}}$} vectors depend only on intra-tile relative coordinates and constant terms, making them independent of specific Gaussians or tile positions. Hence, \textcolor{MyColorBlue}{$\Vec{\bf{v_{p_j}}}$} for all pixels in a tile can be precomputed once offline, stored in global memory, and reused throughout the rendering process. In contrast, \textcolor{MyDeepRed}{$\Vec{\bf{v_{g_i}}}$} must be computed once per Gaussian per tile, as it incorporates both Gaussian parameters and relative offsets $(\hat{x}_{g_i}, \hat{y}_{g_i})$.

This reformulation converts the blending stage into a set of six-dimensional dot products, which can be naturally expressed as matrix multiplications. Specifically, a matrix composed of \textcolor{MyDeepRed}{$\Vec{\bf{v_{g_i}}}$} vectors for multiple Gaussians is multiplied by a matrix of \textcolor{MyColorBlue}{$\Vec{\bf{v_{p_j}}}$} vectors for all pixels in a tile, enabling direct mapping to high-throughput GEMM operations on Tensor Cores.

\begin{algorithm}[tp]
    \caption{GEMM-compatible Blending Process.}
    \label{alg:tensor-blend}
    
    \let\oldnl\nl% Store \nl in \oldnl
    \newcommand{\nonl}{\renewcommand{\nl}{\let\nl\oldnl}}% Remove line number for one line
    
    \KwData{
    Index list of sorted projected Gaussians $\mathcal{G}$ overlapping with current tile, batch size $b$, offline-computed matrix $\textcolor{blue}{\mathbf{M}_p}$
    }
    \KwResult{Rendered Color $\textbf{C}_j$s of pixels in current tile}

    \For{\textnormal{each batch ($b$ Gaussians)} $\subset \mathcal{G}$}
    {
        \BlankLine

        \textcolor[RGB]{113, 152, 107}{/************ \textbf{\texttt{Prepare Data}} *************/}

        Same as Line \textbf{3}-\textbf{8} in Algorithm~\ref{alg:vanilla-blend}

        \BlankLine

        \textcolor[RGB]{113, 152, 107}{/********** \textbf{\texttt{Construct Matrix}} **********/}

        \For{\textnormal{each} Gaussian $\mathcal{G}_i$ in current batch $\textnormal{\textbf{\underline{parallel}}}$ }
        {
            $\Vec{\bf{v_{g_i}}} \leftarrow $ 
            [
                $-\frac{1}{2}A_i$, 
                $-\frac{1}{2}C_i$, 
                $-B_i$, 
                $-A_i \hat{x}_{g_i}-B_i \hat{y}_{g_i}$, 
                $-C_i \hat{y}_{g_i}-B_i \hat{x}_{g_i}$, 
                $-\frac{1}{2}A_i \hat{x}_{g_i}^2$ $-\frac{1}{2}C_i \hat{y}_{g_i}^2$ $-B_i \hat{x}_{g_i}\hat{y}_{g_i}$
            ] \hfill \textcolor[RGB]{88, 100, 113}{// $\texttt{Eq}$~\ref{eq: power rewritten}}
        }

        $\textcolor{MyDeepRed}{\mathbf{M}_g} \leftarrow$
        [
            $\Vec{\bf{v}}_{g_0}^T$,
            $\Vec{\bf{v}}_{g_1}^T$,
            $\dots$,
            $\Vec{\bf{v}}_{g_i}^T$,
            $\dots$,
            $\Vec{\bf{v}}_{g_{b}}^T$
        ]$^T$
        \hfill \textcolor[RGB]{88, 100, 113}{// $\texttt{Eq}$~\ref{eq: vector to matrix}}

        \BlankLine

        % \nonl\textbf{GEMM-Friendly Volume Render:}
        \textcolor[RGB]{113, 152, 107}{/** \textbf{\texttt{GEMM-Compatible Volume Render}} **/}

        $\textcolor{MyBrown}{\mathbf{M}_{power}} \leftarrow \textcolor{MyDeepRed}{\mathbf{M}_g}\cdot \textcolor{MyColorBlue}{\mathbf{M}_p}$
        \hfill
        \textcolor[RGB]{88, 100, 113}{// $\texttt{Eq}$~\ref{eq: matrix multiplication}}

        % \For{$\mathcal{G}_i \in \mathcal{G}$}
        \For{\textnormal{each} Gaussian $\mathcal{G}_i$ in current batch}
        {

            \For{\textnormal{each pixel} $p_j$ \textnormal{in current tile} \textnormal{\textbf{\underline{parallel}}}}
            {

                $\alpha_{ij} \leftarrow o_i \times exp(\textcolor{MyBrown}{\mathbf{M}_{power}}[i][j]) $

                Same as Line \textbf{14}-\textbf{21} in Algorithm~\ref{alg:vanilla-blend}

            }
        }
    }
\end{algorithm}

\begin{figure*}[tp] 
    \setlength{\abovecaptionskip}{3pt}  
    \setlength{\belowcaptionskip}{3pt} 
    \centering
    \includegraphics[width=0.95\linewidth]{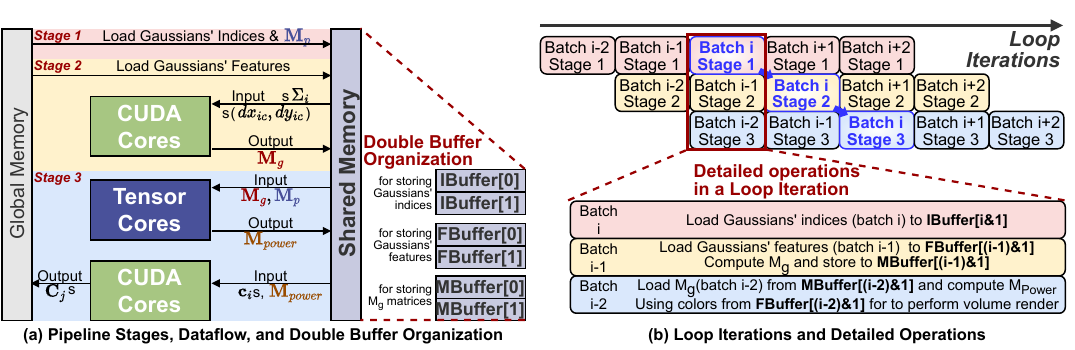}
    \caption{ High-Performance GPU Kernel Design and Implementation. 
        (a) Dataflow of 3-stage pipeline configured with double buffer.
        (b) Execution timing of loop iterations and detailed operations in a single iteration.
    }
    \label{fig: implementation}
    \vspace{-0.3cm}
\end{figure*}

\subsection{GEMM-compatible Blending}

Building on the intra-tile coordinate transformation, we formulate the computation of multiple $power_{ij}$ values as a batched matrix multiplication. This converts the collection of six-dimensional dot products into a single GEMM operation, allowing efficient execution on GPU Tensor Cores for 3DGS acceleration.

For a $16 \times 16$ tile (256 pixels) and a batch of 256 Gaussians---typical in 3DGS rendering~\cite{kerbl20233d}---we construct a $256 \times 6$ Gaussian matrix $\mathbf{M}_g$ and a $256 \times 6$ pixel matrix $\mathbf{M}_p$ based on Eq.~(\ref{eq: power rewritten}):
\begin{equation}
    \label{eq: vector to matrix}
    \begin{aligned}
        &\textcolor{MyDeepRed}{\mathbf{M}_g}
        = \textcolor{MyDeepRed}{
            \begin{bmatrix}
                \Vec{\bf{v}}_{g_0}^T &
                \Vec{\bf{v}}_{g_1}^T &
                \dots &
                \Vec{\bf{v}}_{g_i}^T &
                \dots &
                \Vec{\bf{v}}_{g_{255}}^T
            \end{bmatrix}^T
        }
        \\
        &\textcolor{MyColorBlue}{\mathbf{M}_p}
        = \textcolor{MyColorBlue}{
            \begin{bmatrix}
                \Vec{\bf{v}}_{p_0} & 
                \Vec{\bf{v}}_{p_1} & 
                \dots &
                \Vec{\bf{v}}_{p_i} & 
                \dots &
                \Vec{\bf{v}}_{p_{255}} 
            \end{bmatrix}
        }
    \end{aligned}
\end{equation}

The power values are computed via matrix multiplication:
\begin{equation}
    \label{eq: matrix multiplication}
    \begin{aligned}
        &\textcolor{MyBrown}{\mathbf{M}_{power}} = \textcolor{MyDeepRed}{\mathbf{M}_g}\cdot \textcolor{MyColorBlue}{\mathbf{M}_p}
        \ \ \ \ \ \ \ \ \ \ \ \ \ \ \ \ \ \ \ \ \ \ \ \ \ \ \ 
    \end{aligned}
\end{equation}
where \textcolor{MyBrown}{$\mathbf{M}_{power}$} is a $256 \times 256$ matrix containing all $power_{ij}$ values for Gaussian–pixel pairs within the tile. As discussed in Section~\ref{sec:transf}, \textcolor{MyColorBlue}{$\mathbf{M}_p$} remains identical across tiles due to fixed intra-tile coordinates, enabling offline precomputation and reuse with negligible overhead.

Algorithm~\ref{alg:tensor-blend} summarizes the blending procedure. In contrast to vanilla 3DGS blending (Algorithm~\ref{alg:vanilla-blend}), which evaluates $power_{ij}$ per pixel individually, our method directly consumes the offline-precomputed \textcolor{MyColorBlue}{$\mathbf{M}_p$}. For each tile, the steps are:
1) Fetch Gaussian attributes ($A_i$, $B_i$, $C_i$, $x_{g_i}$, $y_{g_i}$) for the current Gaussian batch.
2) Compute $\textcolor{MyDeepRed}{\Vec{\bf{v_{g_i}}}}$ for each Gaussian $\mathcal{G}i$ using $dx{ic}$ and $\hat{y}_{g_i}$, forming \textcolor{MyDeepRed}{$\mathbf{M}_g$}.
3) Perform matrix multiplication (Eq.(\ref{eq: matrix multiplication})) to obtain \textcolor{MyBrown}{$\mathbf{M}_{power}$}.
4) Execute volume rendering, computing $\alpha_{ij}$ using exponentials of \textcolor{MyBrown}{$\mathbf{M}_{power}$} values, consistent with Algorithm~\ref{alg:vanilla-blend}.

\subsection{GPU Implementation}

Following the vanilla implementation~\cite{kerbl20233d}, we assign a thread block of 256 threads to render each $16 \times 16$ tile and process the sorted Gaussian list with a batch size of $256$.

\textbf{Three-Stage Pipeline.}
To overlap latency and reduce synchronization overhead, we design a three-stage pipeline, illustrated in Figure~\ref{fig: implementation}(a):
\underline{Stage 1}: Load Gaussian indices into shared memory.
\underline{Stage 2}: Each thread fetches the attributes of one Gaussian according to its index (as in line~4 of Algorithm~\ref{alg:tensor-blend}), and collaboratively constructs $\mathbf{M}_g$. Each thread computes one \textcolor{MyDeepRed}{$\Vec{v}{g_i}$} vector, forming the rows of $\mathbf{M}_g$.
\underline{Stage 3}: Perform matrix multiplication between the preloaded \textcolor{MyColorBlue}{$\mathbf{M}_p$} and the constructed \textcolor{MyDeepRed}{$\mathbf{M}_g$} to obtain \textcolor{MyBrown}{$\mathbf{M}_{power}$}, followed by volume rendering.

\begin{table}[tp]
    \centering
    \setlength{\abovecaptionskip}{2pt}  
    \setlength{\belowcaptionskip}{2pt} 
    \caption{3DGS Workloads Statics.}
    \resizebox{1.0\linewidth}{!}{
    % \addtolength{\tabcolsep}{-2pt}
    \begin{tabular}{cccc}
        \Xhline{1.5px}
        \textbf{Datasets} & \textbf{Scenes} & \textbf{Resolution} & \textbf{\#Gaussians} \\ 
        \Xhline{1.5px}
        \multirow{2}{*}{\begin{tabular}[c]{@{}c@{}}Tank \& Temples\\ (Outdoor)~\cite{knapitsch2017tanks}\end{tabular}} & Train & $980\times 545$ & 1.09M \\ \cline{2-4} 
         & Truck & $979\times 546$ & 2.06M \\ \hline
        \multirow{2}{*}{\begin{tabular}[c]{@{}c@{}}Deep Blending\\ (Indoor)~\cite{hedman2018deep}\end{tabular}} & Playroom & $1264\times 832$ & 1.85M \\ \cline{2-4} 
         & Drjohnson & $1332\times 876$ & 3.07M \\ \hline
        \begin{tabular}[c]{@{}c@{}}Mip-NeRF 360\\ (5 outdoors,\\ 4 indoors)~\cite{barron2022mip}\end{tabular} & \begin{tabular}[c]{@{}c@{}}bicycle, bonsai,\\ counter, flowers,\\ garden, kitchen,\\ room, stump, treehill\end{tabular} & \begin{tabular}[c]{@{}c@{}} around \\ $1060\times 1600$ \end{tabular} & \begin{tabular}[c]{@{}c@{}} $1.04$M  $\sim$ \\ $4.74$M \end{tabular} \\ % \textasciitilde
        \Xhline{1.5px}
        \end{tabular}
    }
    \label{tab: 3dgs datasets}
    \vspace{-0.3cm}
\end{table}
\textbf{Mini-batch Design with PTX-based GEMM Instruction.}
Due to limited shared memory, Stage~3 adopts a mini-batch design. 
In each iteration, a $16 \times 8$ submatrix of \textcolor{MyDeepRed}{$\mathbf{M}_g$} is multiplied with \textcolor{MyColorBlue}{$\mathbf{M}_p$}. % modified by lhm, to check
The multiplication is executed via four rounds of warp-level \texttt{mma(m16n8k8)} PTX instructions\footnote{\url{https://docs.nvidia.com/cuda/parallel-thread-execution/index.html\#warp-level-matrix-fragment-mma-1688}.}
, where each thread block consists of 8 warps (32 threads per warp). Thus, $8 \times 4 = 32$ warp-level calls complete an effective \texttt{m256n16k8} multiplication. This partitioning confines synchronization to the warp level, eliminating costly thread block barriers.

\textbf{Double Buffer Organization.}
We employ double buffering for Gaussian indices, features, and the matrix \textcolor{MyDeepRed}{$\mathbf{M}_g$}, as illustrated in Figure~\ref{fig: implementation}(b).
Double buffering of indices enables the use of asynchronous memory copy\footnote{\url{https://docs.nvidia.com/cuda/cuda-c-programming-guide/index.html\#pipeline-primitives-interface}.}
, introduced in the Ampere architecture, to overlap data transfer with computation. Similarly, double buffering of features and \textcolor{MyDeepRed}{$\mathbf{M}_g$} prevents synchronization stalls between Stage~2 and Stage~3. This design minimizes global memory traffic, ensuring sustained tensor-core utilization.

\begin{table*}[tp]
    \setlength{\abovecaptionskip}{2pt}
    \setlength{\belowcaptionskip}{2pt}
    \caption{ Average image rendering latency (\textit{ms}) comparison between GEMM-GS and baseline methods on an A100 GPU.}
    \label{tab: a100 results}
    \resizebox{1\linewidth}{!}{
    \addtolength{\tabcolsep}{-2pt}
    \begin{tabular}{cccccccccccccc}
    \Xhline{2px}
    \multicolumn{1}{c}{\textbf{Datasets}} & \multicolumn{2}{|c}{\textbf{Tank \& Temples~\cite{knapitsch2017tanks}}} & \multicolumn{2}{|c}{\textbf{Deep Blending~\cite{hedman2018deep}}} & \multicolumn{9}{|c}{\textbf{Mip-NeRF 360~\cite{barron2022mip}}} \\ \Xhline{1.2px}
    \multicolumn{1}{c}{\diagbox{\textbf{Methods}}{\textbf{Scenes}}} & \textbf{truck} & \textbf{train} & \textbf{drjohnson} & \textbf{playroom} & \textbf{bicycle} & \textbf{bonsai} & \textbf{counter} & \textbf{flowers} & \textbf{garden} & \textbf{kitchen} & \textbf{room} & \textbf{stump} & \textbf{treehill} \\ \Xhline{1.5px}
    Vanilla 3DGS & 4.28 & 4.51 & 6.16 & 4.37 & 9.64 & 3.77 & 5.17 & 4.22 & 6.49 & 5.97 & 5.32 & 4.73 & 5.02 \\ \hline
    \ \textbf{+ GEMM-GS} & 2.78 & 3.01 & 4.16 & 2.86 & 6.87 & 2.7 & 3.72 & 2.99 & 4.81 & 4.33 & 3.97 & 3.47 & 3.63 \\ \hline
    \ \ \textbf{Speedup} & \textbf{1.54$\times$} & \textbf{1.50$\times$} & \textbf{1.48$\times$} & \textbf{1.53$\times$} & \textbf{1.40$\times$} & \textbf{1.40$\times$} & \textbf{1.39$\times$} & \textbf{1.41$\times$} & \textbf{1.35$\times$} & \textbf{1.38$\times$} & \textbf{1.34$\times$} & \textbf{1.36$\times$} & \textbf{1.38$\times$} \\ \Xhline{1.5px}
    FlashGS & 2.53 & 2.6 & 2.08 & 1.51 & 4.16 & 1.52 & 2.08 & 2.31 & 3.74 & 2.84 & 2.02 & 2.71 & 2.48  \\ \hline
    \ \textbf{+ GEMM-GS} & 2.29 & 2.15 & 1.58 & 1.23 & 3.69 & 1.27 & 1.73 & 1.93 & 3.21 & 2.39 & 1.72 & 2.28 & 2.15 \\ \hline
    \ \ \textbf{Speedup} & \textbf{1.10$\times$}  & \textbf{1.21$\times$}  & \textbf{1.32$\times$}  & \textbf{1.23$\times$}  & \textbf{1.13$\times$}  & \textbf{1.20$\times$}  & \textbf{1.20$\times$}  & \textbf{1.20$\times$}  & \textbf{1.17$\times$}  & \textbf{1.19$\times$}  & \textbf{1.17$\times$}  & \textbf{1.19$\times$}  & \textbf{1.15$\times$} \\  
    \Xhline{1.5px}
    StopThePop & 3.52 & 3.86 & 5.49 & 3.53 & 8.91 & 2.91 & 4.31 & 3.62 & 5.74 & 5.31 & 4.72 & 4.12 & 4.28 \\ \hline
    \ \textbf{+ GEMM-GS} & 2.56 & 2.67 & 3.75 & 2.23 & 6.28 & 2.016 & 3.11 & 2.63 & 4.324 & 3.83 & 3.35 & 2.928 & 3.048 \\ \hline
    \ \ \textbf{Speedup} & \textbf{1.38$\times$} & \textbf{1.45$\times$} & \textbf{1.46$\times$} & \textbf{1.58$\times$} & \textbf{1.42$\times$} & \textbf{1.45$\times$} & \textbf{1.39$\times$} & \textbf{1.38$\times$} & \textbf{1.33$\times$} & \textbf{1.39$\times$} & \textbf{1.41$\times$} & \textbf{1.41$\times$} & \textbf{1.40$\times$}  \\  
    \Xhline{1.5px}
    Speedy-Splat & 2.83 & 2.93 & 2.85 & 2.01 & 4.7 & 1.91 & 2.5 & 2.61 & 4.16 & 3.37 & 2.31 & 2.99 & 2.82 \\ \hline
    \ \textbf{+ GEMM-GS} & 1.86 & 1.88 & 1.79 & 1.26 & 3.38 & 1.25 & 1.59 & 1.78 & 2.79 & 2.44 & 1.44 & 2.09 & 2.04 \\ \hline
    \ \ \textbf{Speedup} & \textbf{1.52$\times$}  & \textbf{1.56$\times$}  & \textbf{1.59$\times$}  & \textbf{1.60$\times$}  & \textbf{1.39$\times$}  & \textbf{1.53$\times$}  & \textbf{1.57$\times$}  & \textbf{1.47$\times$}  & \textbf{1.49$\times$}  & \textbf{1.38$\times$}  & \textbf{1.60$\times$}  & \textbf{1.43$\times$}  & \textbf{1.38$\times$}  \\  
    \Xhline{1.5px}
    c3dgs & 4.3 & 4.93 & 6.22 & 4.35 & 10.85 & 4.37 & 5.92 & 4.71 & 7.83 & 7.37 & 6 & 5.53 & 6.09 \\ \hline
    \ \textbf{+ GEMM-GS} & 2.53 & 2.89 & 3.49 & 2.44 & 6.29 & 2.55 & 3.4 & 2.87 & 4.56 & 4.18 & 3.37 & 3.18 & 3.61 \\ \hline
    \ \ \textbf{Speedup} & \textbf{1.70$\times$} & \textbf{1.71$\times$} & \textbf{1.78$\times$} & \textbf{1.78$\times$} & \textbf{1.72$\times$} & \textbf{1.71$\times$} & \textbf{1.74$\times$} & \textbf{1.64$\times$} & \textbf{1.72$\times$} & \textbf{1.76$\times$} & \textbf{1.78$\times$} & \textbf{1.74$\times$} & \textbf{1.69$\times$}   \\  
    \Xhline{1.5px}
    LightGaussian & 2.9 & 2.29 & 3.55 & 2.7 & 5.96 & 2.95 & 3.94 & 3.38 & 5.14 & 4.17 & 4 & 3.85 & 4.18 \\ \hline
    \ \textbf{+ GEMM-GS} & 1.85 & 1.51 & 2.2 & 1.68 & 3.67 & 1.88 & 2.48 & 2.18 & 3.23 & 2.62 & 2.49 & 2.47 & 2.6 \\ \hline
    \ \ \textbf{Speedup} & \textbf{1.57$\times$} & \textbf{1.52$\times$} & \textbf{1.61$\times$} & \textbf{1.61$\times$} & \textbf{1.62$\times$} & \textbf{1.57$\times$} & \textbf{1.59$\times$} & \textbf{1.55$\times$} & \textbf{1.59$\times$} & \textbf{1.59$\times$} & \textbf{1.61$\times$} & \textbf{1.56$\times$} & \textbf{1.61$\times$}   \\  
    \Xhline{2px}
    \end{tabular}  
    }
    \vspace{-0.3cm}
\end{table*}

\begin{figure*}[tp] 
    \setlength{\abovecaptionskip}{1pt}  
    \setlength{\belowcaptionskip}{1pt} 
    \centering
    \includegraphics[width=1.0\linewidth]{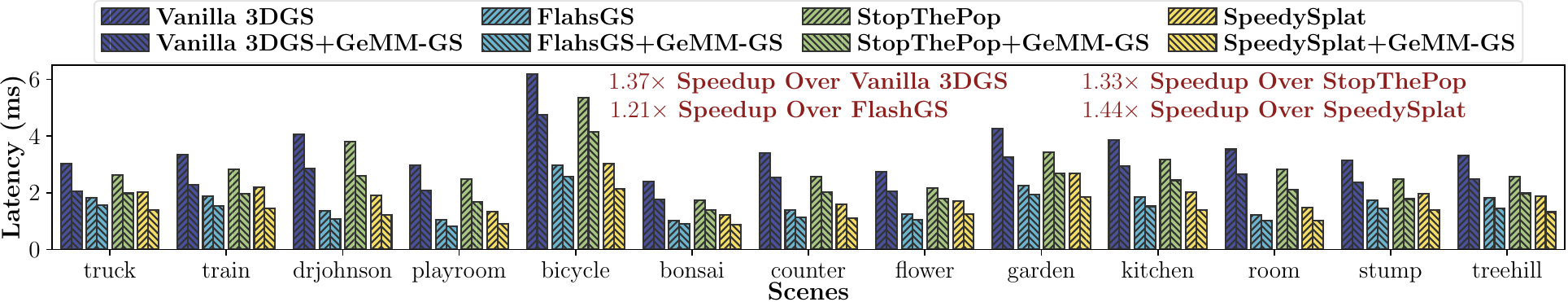}
    \caption{ Average image rendering latency (\textit{ms}) comparison between GEMM-GS and baseline methods on an H100 GPU.
    }
    \label{exp-fig: h100}
    \vspace{-0.4cm}
\end{figure*}

\section{Evaluation}

\subsection{Methodology}

\textbf{Workloads and Datasets.}
We evaluate 3DGS on three widely-used real-world datasets: Tank \& Temples~\cite{knapitsch2017tanks}, Deep Blending~\cite{hedman2018deep}, and Mip-NeRF 360~\cite{barron2022mip}. Each scene is trained for 30K iterations using the official 3DGS pipeline. In total, we evaluate across 13 scenes (workloads), with dataset statistics summarized in Table~\ref{tab: 3dgs datasets}.

\textbf{Implementation and Baselines.}
We implement our design on 2 types of NVIDIA GPUs: A100 and H100.
We use CUDA events to delay the forward rendering process within the forward function. For each scene, we iterate through rendering the entire test dateset 10 times, record the latency, and take the average.
Apart from the Vanilla 3DGS implementation~\cite{kerbl20233d}, we select several representive 3DGS accleration as baselines.
\underline{Preprocess Optimization Methods:}
FlashGS~\cite{feng2025flashgs}, StopThePop~\cite{radl2024stopthepop}, and Speedy-Splat~\cite{hanson2025speedy}.
\underline{Compression Methods:}
LightGaussian~\cite{fan2024lightgaussian} and c3dgs~\cite{lee2024compact}.
We compare the latency of these baseline designs with the latency after integrating the proposed GEMM-GS into these designs.

\subsection{Performance Comparison}

Table~\ref{tab: a100 results} and Figure~\ref{exp-fig: h100} compares the latency between baseline methods and our GEMM-GS.
Compared to the vanilla 3DGS, GEMM-GS achieves $1.42\times$ and $1.37\times$ speedup on A100 and H100, respectively. Such improvements are due to utilization of powerful tensor cores on GPUs and efficient pipeline kernel design.
Furthermore, we integrate GEMM-GS with several preprocess optimization methods.
GEMM-GS provides averagely $1.19\times$ ($1.21\times$), $1.42\times$ ($1.33\times$), and $1.50\times$ ($1.44\times$) over FlashGS, StopThePop, and SpeedySplat on A100 (H100), demonstrating that our method is orthogonal to existing methods and can be combined with them to achieve co-optimization for better performance.
In addition, we evaluate GEMM-GS integrated with compression methods on A100.
GEMM-GS achieves $1.73\times$ and $1.58\times$  over c3dgs and LightGaussian on average, indicating that our method is also friendly to compression methods that pruning redundant computation.

\subsection{Sensitivity study}

\begin{figure}[tp] 
    \setlength{\abovecaptionskip}{1pt}  
    \setlength{\belowcaptionskip}{1pt} 
    \centering
    \includegraphics[width=1.0\linewidth]{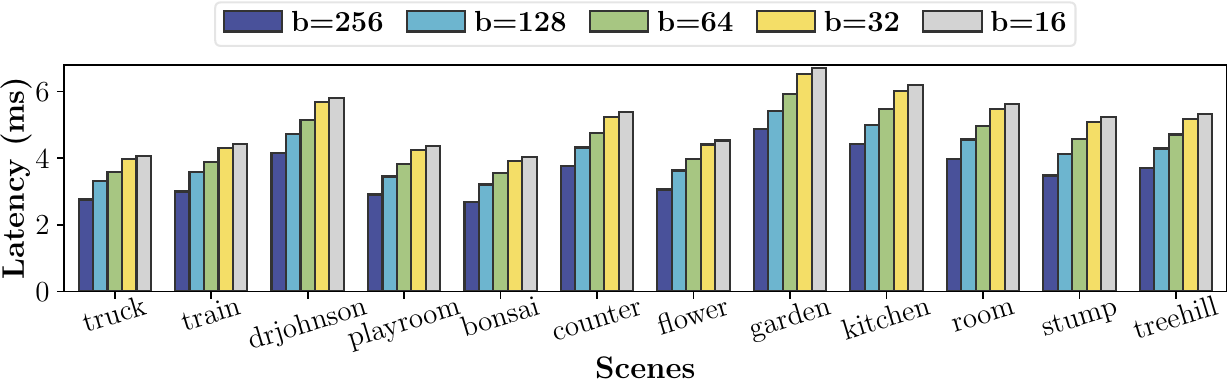}
    \caption{ Average rendering latency (\textit{ms}) comparison between GEMM-GS and vanilla 3DGS with various batch size $b$.
    }
    \label{exp-fig: sensitivity batchsize}
    \vspace{-0.3cm}
\end{figure}

\textbf{Impact of Batch Size.}
Figure~\ref{exp-fig: sensitivity batchsize} presents the latency comparison under different batch sizes. Smaller batches result in higher latency due to poor parallel acceleration.
Specifically, a tile (16$\times$16 pixels) is allocated 256 threads, so when the batch size is smaller than 256, the task of constructing matrix $\mathbf{M}_g$ cannot be evenly divided into each thread, and smaller workload causes worse parallel acceleration.

\begin{figure}[tp] 
    \setlength{\abovecaptionskip}{1pt}  
    \setlength{\belowcaptionskip}{1pt} 
    \centering
    \includegraphics[width=1.0\linewidth]{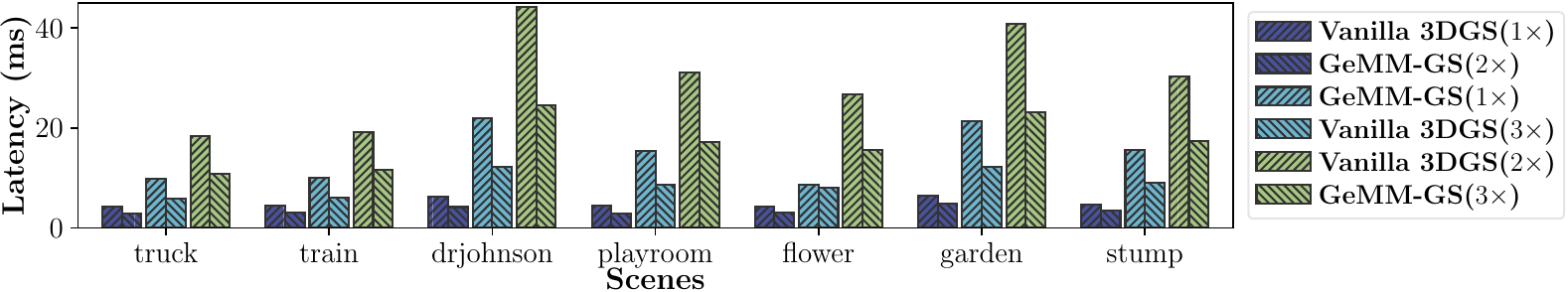}
    \caption{ Average image rendering latency (\textit{ms}) comparison between GEMM-GS and vanilla 3DGS with various resolution ($1\times$, $2\times$, $3\times$).
    }
    \label{exp-fig: sensitivity resolution}
    \vspace{-0.3cm}
\end{figure}

\textbf{Impact of Resolution.}
We evaluate GEMM-GS on rendering images with various resolutions, from $1\times$ to $3\times$, as shown in Figure~\ref{exp-fig: sensitivity resolution}.
GeMM-GS provides $1.73\times$ and $1.74\times$ speedup over vanilla 3DGS on $2\times$ and $3\times$ resolution, emphasizing the scalability of our method on high-resolution scenarios.

\section{Conclusion}

This paper presents GEMM-GS, an approach to accelerating 3DGS rendering on modern GPUs with tensor cores.
By leveraging intra-tile relative coordinates, we propose a GEMM-compatible variant of the 3DGS blending operation that enables acceleration via tensor cores.
While TC-GS \cite{liao2025tc} explores Tensor Core integration for 3DGS, GEMM-GS distinguishes itself through architecture-aware optimizations, including asynchronous data fetch and mini-batch design with PTX-based GEMM instructions. By maximizing GPU hardware utilization, GEMM-GS achieves superior speedups over existing acceleration methods.

\bibliographystyle{ACM-Reference-Format}
\bibliography{ref}

\end{document}